\documentclass[aps,prl,twocolumn,superscriptaddress,longbibliography]{revtex4-2}

\usepackage{mathrsfs}
\usepackage{tabularx}
\usepackage{slashed}
\usepackage{amsmath}
\usepackage{amsfonts}
\usepackage{bm}
\usepackage{graphicx,color}
\usepackage{times}
\usepackage{braket}
\usepackage{orcidlink}
\usepackage[caption=false]{subfig}
\AtBeginDocument{%
   ~\newwrite\bibnotes
   ~\def\bibnotesext{Notes.bib}
   ~\immediate\openout\bibnotes=\jobname\bibnotesext
   ~\immediate\write\bibnotes{@CONTROL{REVTEX41Control}}
   ~\immediate\write\bibnotes{@CONTROL{%
    apsrev42Control,author="08",editor="1",pages="1",title="0",year="1"}}
    ~\if@filesw
    ~\immediate\write\@auxout{\string\citation{apsrev42Control}}%
   ~\fi
}%

\begin{document}
\title{Topological Order Without Band Topology in Moiré Graphene}

\author{Hui Liu \orcidlink{0009-0009-4988-9561}}\thanks{hui.liu@fysik.su.se}
\affiliation{Department of Physics, Stockholm University, AlbaNova University Center, 106 91 Stockholm, Sweden}
\author{Raul Perea-Causin  \orcidlink{0000-0002-2229-0147}}\thanks{raul.perea.causin@fysik.su.se}
\affiliation{Department of Physics, Stockholm University, AlbaNova University Center, 106 91 Stockholm, Sweden}
\author{Zhao Liu \orcidlink{0000-0002-3947-4882}}\thanks{zhaol@zju.edu.cn}
\affiliation{Zhejiang Institute of Modern Physics, Zhejiang University, Hangzhou 310058, China}
\affiliation{Zhejiang Key Laboratory of Micro-Nano Quantum Chips and Quantum Control, School of Physics, Zhejiang University, Hangzhou 310027, China}
\author{Emil J. Bergholtz \orcidlink{0000-0002-9739-2930}}\thanks{emil.bergholtz@fysik.su.se}
\affiliation{Department of Physics, Stockholm University, AlbaNova University Center, 106 91 Stockholm, Sweden}

\date{\today}

\begin{abstract}
The discovery of zero-field fractional Chern insulators (FCIs) in moiré materials has attracted intense interest in the interplay between topology and correlations. 
Here, we demonstrate that fractionalized topological order can emerge under realistic conditions even within a topologically trivial moiré band. 
Employing exact diagonalization of
long-range Coulomb interactions projected into a trivial band of twisted multilayer graphene, we identify a set of incompressible Laughlin-like ground states exhibiting fractional quantized Hall conductance. 
We trace the origin of this FCI phase to the strongly inhomogeneous distribution of quantum geometry, which reshapes interaction effects independently of the band topology.
Extending this heuristic quantum geometric mechanism, we demonstrate that similarly unexpected Laughlin-like FCIs can also be stabilized in higher-Chern-number moiré bands under experimentally accessible conditions. 
Our results establish realistic scenarios under which many-body topological order can emerge independently of single-particle band topology.
\end{abstract}
\maketitle
The emergence of topological flat bands in highly tunable moiré superlattices has established twisted van der Waals stacks as an ideal platform for exploring correlated topological phases~\cite{andrei2020graphene,andrei2021marvels,mak2022semiconductor}. 
A landmark achievement in this area is the observation of zero-field fractional Chern insulators (FCIs)~\cite{xie2021fractional,FCI_MoTe2_3,FCI_MoTe2_1,FCI_MoTe2_2,lu2024fractional,PhysRevX.13.031037}. 
As lattice analogues of the fractional quantum Hall effect, FCIs exhibit dissipationless quantized Hall conductance and host fractionalized anyonic quasiparticles at relatively high temperatures, garnering intense interest for both fundamental understanding and technological applications of quantum materials~\cite{spanton2018observation,high_temperature_FCI,PhysRevX.1.021014,LIU2024515, bergholtz2013topological,PARAMESWARAN2013816,kolread,tang_high-temperature_2011,PhysRevLett.106.236803,PhysRevLett.106.236804,sheng2011fractional,mollercooper,PhysRevLett.105.215303,PhysRevLett.124.106803,repellinChernBandsTwisted2020,PhysRevResearch.2.023237,ZhaoTDBG,PhysRevResearch.3.L032070,PhysRevB.103.125406}.

Following this milestone, subsequent research has unveiled even more exotic correlated phases. Among these are the Moore-Read state at half filling~\cite{mr_Aidan,mr_Wang,mr_Xu,mr_Ahn,mr_Hui,mr_Donna,mr_zhao,mr_sen} and the Read-Rezayi state at $\nu=3/5$ filling~\cite{liu2024parafermions}, with the former showing possible experimental signatures~~\cite{kang2024evidence,HiddenFCI}. These phases host non-Abelian anyons as their elementary excitations, and thus constitute a potential platform for fault-tolerant quantum computation. 
In parallel, the quantum (anomalous) Hall crystal has been predicted and realized in moiré materials. This phase breaks either continuous or discrete translation symmetry and exhibits a surprising coexistence of topology and crystalline order at fractional fillings, phenomena that are traditionally considered to be mutually exclusive~\cite{tesanovic1989hall,song2024intertwined,ahc2024,kwan2023moirefractionalcherninsulators,PhysRevLett.133.206502,dong2024anomalous,dong2024stability,zhou2024fractionalquantumanomaloushall,soejima2024anomalous,PhysRevX.14.041040,paul2024designinghigherhallcrystals,zhou2024newclassesquantumanomalous,PhysRevLett.132.236601,lu2024extended,polshyn2022topological,su2024topological,waters2024interplay}. 
More recently, correlated topological states have been explored in moiré bands with higher Chern number ($|C|>1$), a regime accessible in moiré superlattices that transcends the conventional Landau-level paradigm~\cite{su2024topological,fci_zhao_hc, fci_Sterdyniak_hc, fci_Behrmann_hc,Raul_hall_crystal,sen_higher_chern_band_hall_crystal,Choi_higher_chern}. A key development has been the observation of both higher-Chern-number quantum anomalous Hall crystals and FCIs in twisted multi-layer graphene~\cite{higher_chern_band_FCI1,higher_chern_band_FCI2,higher_chern_band_FCI3}.
These discoveries highlight the rich interplay of band topology, interactions, and symmetry breaking in moiré systems.

While all these advances have relied on nontrivial band topology, theoretical studies have shown that quantum Hall-like physics can emerge even in a single topologically trivial band~\cite{hui_topological_fine,Jan_quantum_hall_without_chern}.
This type of trivial band topology is fundamentally different from the zero Hall conductance of a pair of time-reversal conjugated Chern bands: quantum Hall physics is naturally expected in the latter case if spontaneous polarization to one of the two bands is present, but it is counterintuitive in the former.
Examples of models supporting FCIs in a single trivial band have thus far been exquisitely designed to make the conceptual point: either by artificially embedding topological charges into many-body interactions~\cite{Simon15} or by requiring short-range interactions in toy models~\cite{Lin_zero_chern_FCI}.
Whether trivial-band FCIs could be stabilized by realistic long-range Coulomb interactions, especially in experimentally accessible materials, has remained an open question.

Here, we demonstrate that moiré bands with trivial topology can host robust FCIs under realistic interactions. In particular, we identify Laughlin-like states with fractionally quantized Hall conductance emerging from the fractional filling of a topologically trivial ($C=0$) moiré band with strongly inhomogeneous Berry curvature and Fubini-Study (FS) metric.
In addition, with the same geometric mechanism we resolve an interesting puzzle in twisted double bilayer graphene (TDBG), which hosts a nearly flat $C=2$ conduction band. At electron band filling $\nu=1/3$, we observe an FCI with unanticipated Hall conductance $\sigma_{xy}=1/3$ (in units of $e^2/h$), deviating from the conventional expectation of $\sigma_{xy}=2/3$. Many-body diagnostics confirm its topological character, which originates from the same geometric mechanism operative in the trivial-band case.
Our findings establish the distribution of quantum geometry---rather than the band Chern number---as the key aspect governing the formation of FCIs, and pose moiré systems as a promising experimental platform for realizing these physics.

\begin{figure}[t]
\centering
\includegraphics[width=\linewidth]{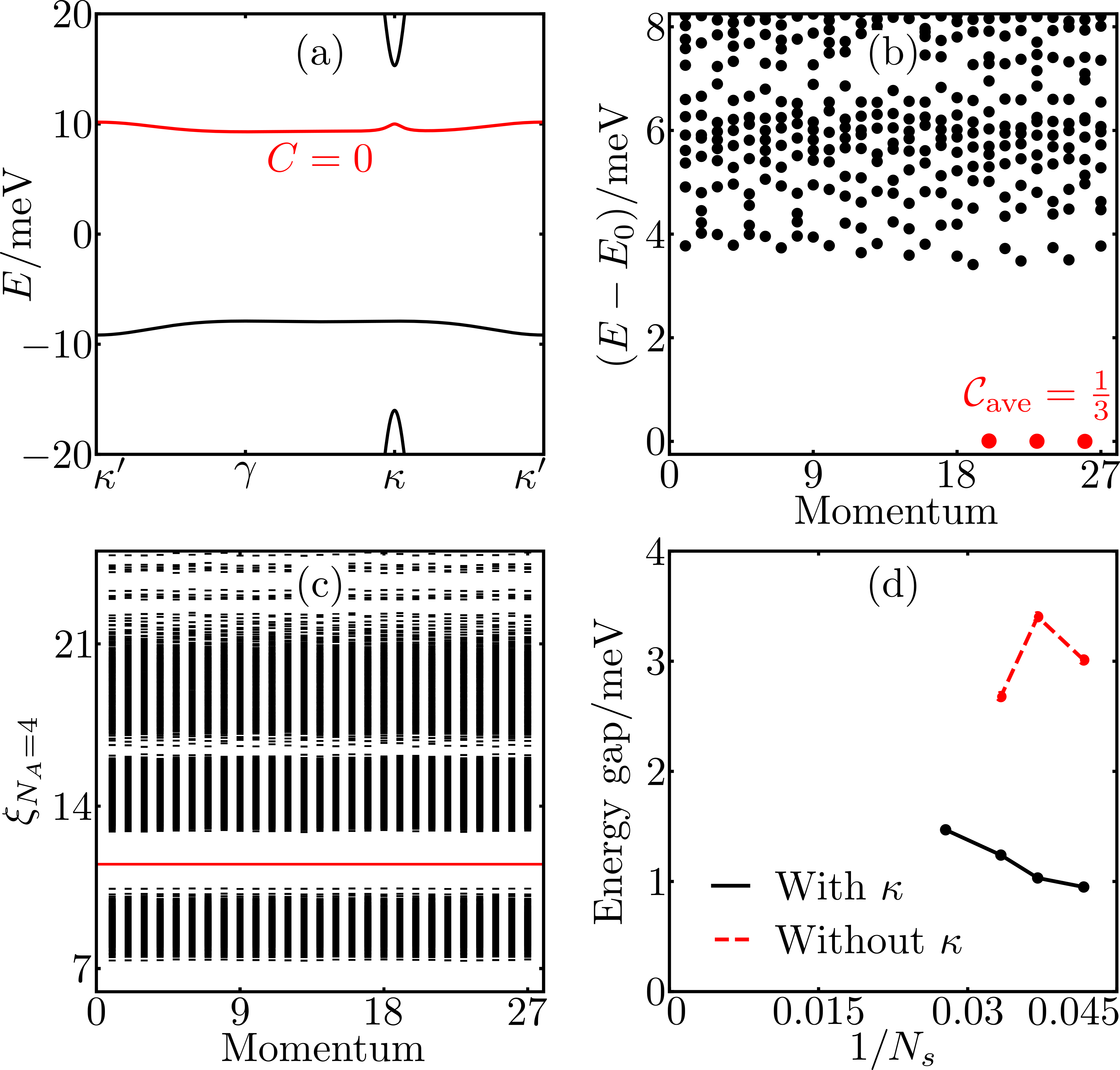}
\caption{\textbf{FCIs in a \boldmath{$C=0$ band}}. (a) Single-particle band structure displaying a topologically trivial conduction band (marked in red). (b) Low-lying many-body energy spectrum, (c) PES, and (d) energy gap scaling demonstrating the stabilization of Laughlin-like FCIs at $1/3$ filling of the trivial band.
The data shown in (b) and (c) correspond to a 27-site system 
which excludes the $\kappa$ point from the set of discrete accessible momenta.
The number of states below the solid line in (c) matches the quasihole counting of the $1/3$ Laughlin state. Panel (d) displays the many-body energy gap as a function of the system size. Samples in which the $\kappa$ point is accessible are also considered. Additional details, including the system parameters, can be found in the SM~\cite{SupMat}, which includes Ref.~\cite{PhysRevB.90.245401}.
}
\label{fig:1}
\end{figure}

\emph{Setup}---%
We consider twisted multilayer graphene, which can host a nearly flat conduction band with tunable Chern number that is well separated from remote bands by sizable energy gaps~\cite{PhysRevLett.128.176404,jie_wang_qgt1,ZhaoTDBG}. 
The many-body Hamiltonian projected to this target band reads
\begin{eqnarray}
    H=\sum_{\bf k}\epsilon_{\bf k}c^\dagger_{\bf k}c_{\bf k}+ \frac{1}{2A} \sum_{\{\mathbf{k}_i\}}V_{\mathbf{k}_1\mathbf{k}_2\mathbf{k}_3\mathbf{k}_4}c^\dagger_{\mathbf{k}_1}c^\dagger_{\mathbf{k}_2}c_{\mathbf{k}_3}c_{\mathbf{k}_4},
\end{eqnarray}
where $\epsilon_{\bf k}$ is the electron's band dispersion, $c^\dagger_{\mathbf{k}}$ ($c_{\mathbf{k}}$) denotes the electron creation (annihilation) operator at momentum $\mathbf{k}$, $A$ is the area of the lattice, and $V_{\mathbf{k}_1\mathbf{k}_2\mathbf{k}_3\mathbf{k}_4}$ is the interaction matrix element, which contains single-particle wavefunction overlaps and enforces momentum conservation. 
Here, we consider the bare Coulomb interaction $V(\mathbf{q})=\frac{e^2}{2\epsilon\epsilon_0|\mathbf{q}|}$ with dielectric constant $\epsilon\approx 4$, which is typically used for describing interactions in graphene systems~\cite{jung2019}. 

We apply exact diagonalization to finite systems of $N_e$ electrons and $N_s$ moir\'e unit cells. The spanning vectors defining the finite systems considered are detailed in the SM~\cite{SupMat}. To characterize the ground state, we extract a suite of key observables: the low-lying energy spectrum, particle-cut entanglement spectrum (PES)~\cite{PhysRevX.1.021014,LIU2024515}, structure factor, and many-body Chern number. This multifaceted approach is essential for reliably differentiating FCIs from competing orders such as charge density waves (CDWs), quantum Hall crystals, and Fermi liquids. 

\begin{figure}[t]
\centering
\includegraphics[width=\linewidth]{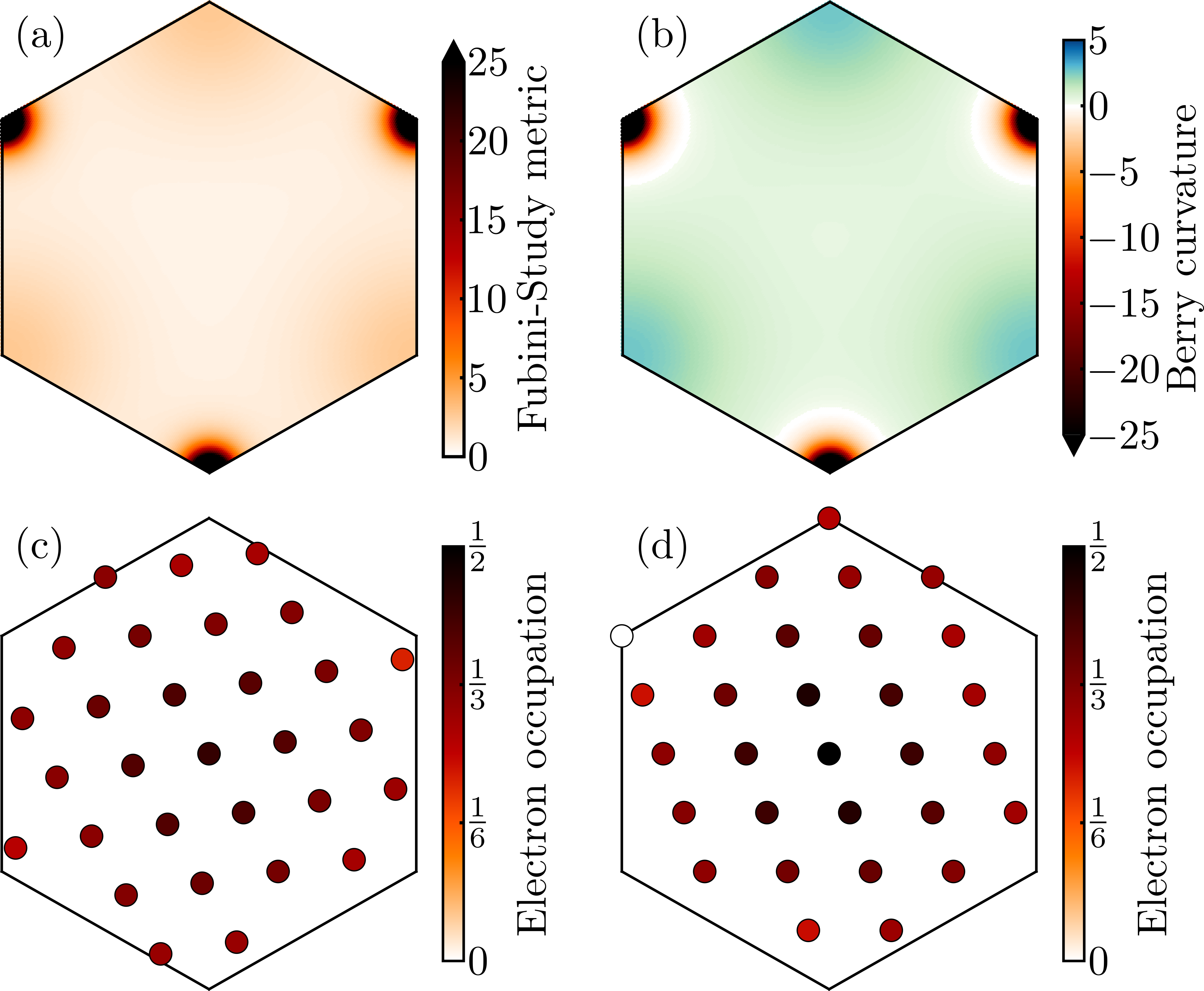}
\caption{\textbf{Quantum geometry perspective on trivial-band FCIs}. (a) Trace of the FS metric and (b) Berry curvature of the trivial band displaying a clear concentration around the $\kappa$ point. (c) The electron occupation is uniformly distributed along the mBZ in finite systems excluding the high-symmetry point $\kappa$. (d) In finite systems containing the $\kappa$ point, the electron occupation remains uniform but exhibits a hole at the $\kappa$ point. This behavior appears consistently for different system sizes (see SM~\cite{SupMat}).
}
\label{fig:2}
\end{figure}

\emph{FCIs in trivial moiré bands}---%
We first consider a trivial conduction band in small-angle twisted bilayer graphene (TBG). In the case of polarization (i.e., broken time-reversal symmetry) into the $K$ valley, the moiré conduction band carries Chern number $C=1$. However, a key mechanism can give rise to the trivial band of interest for our work: the weak coupling between the $K$ and $K'$ valleys of the bottom graphene layer.
Such a scenario can be experimentally realized by aligning TBG with a commensurate insulating substrate that generates an intervalley coupling within the bottom graphene layer~\cite{intervalley_coupling1,intervalley_coupling}. Through this process, the opposite chirality from the $K'$ valley of the bottom layer cancels the Berry curvature contribution from the $K$ valley of the same layer, thereby annihilating the net Chern number of the conduction band. 
In particular we observe that, while the valley polarized TBG is characterized by a nearly uniform positive Berry curvature, the weak intervalley coupling leads to a highly-concentrated negative Berry curvature at one of the inequivalent corners of the moiré Brillouin zone (mBZ); see Fig.~\ref{fig:2}(b).
As a result of the exact cancellation between the regions with positive and negative Berry curvature, the originally topological $C=1$ band is transformed into an isolated and fully trivial $C=0$ band; see Fig.~\ref{fig:1}(a).
Thus, the system at filling $\nu=1$ is, in principle, a trivial insulator---although its nature can be different due to interaction-induced band mixing.

After identifying the underlying mechanism for realizing the trivial band of interest, we explore the nature of correlated states emerging at fractional filling of this band, focusing on $\nu=1/3$. 
We find that the low-lying energy spectrum of the many-body Hamiltonian exhibits a characteristic pattern that remains consistent across various system sizes (see Supplemental Material, SM~\cite{SupMat}). In particular, the ground states are threefold degenerate, their center-of-mass momenta are consistent with the Haldane statistics of the $\nu=1/3$ Laughlin state~\cite{Bergholtz2008}, and they are separated from excited states by a clear gap; c.f. Fig.~\ref{fig:1}(b). 
To characterize the topological nature of these states, we compute their many-body Chern numbers---which are defined in terms of the Berry curvature of many-body states in the parameter space of inserted magnetic flux (twisted boundary conditions)~\cite{Niu_manybodyChern,Kudo_manybodyChern}.
The resulting many-body Berry curvature is remarkably smooth, indicating the absence of any direct gap closing (see SM~\cite{SupMat}).
The many-body Chern numbers surprisingly reveal that each ground state shares a non-zero Hall conductance $\sigma_{xy}=1/3$.
This finding is in stark contrast to conventional trivial flat bands where the only correlated phases are expected to be trivial CDWs with $\sigma_{xy}=0$. 

To further exclude the possibility of CDW order, we turn to the PES analysis~\cite{PhysRevX.1.021014,LIU2024515}.
Here, the system is divided into subsystems $A$ and $B$ with $N_A$ and $N_e-N_A$ particles, respectively.
The PES consists of the eigenvalues $\{\xi\}$ of $-\text{log}\rho_A$, where $\rho_A=\text{tr}_B[\frac{1}{N_d}\sum_{i=1}^{N_d}|\psi_i\rangle\langle \psi_i|]$ is the reduced density matrix of subsystem $A$, $N_d$ is the ground-state degeneracy and $|\psi_i\rangle$ denotes the $i$-th ground state.
A low value of $\xi$ indicates a large probability for a specific quasihole excitation to occur---thus, the PES reveals the unique statistics of quasihole excitations.
As shown in Fig.~\ref{fig:1}(c) and the SM~\cite{SupMat}, the PES shows a large number of states below the entanglement gap---indicative of a highly-entangled phase instead of a CDW. In fact, the number of states below the entanglement gap matches exactly the analytical counting of quasihole excitations in $1/3$ Laughlin states, fully supporting the identification of the observed many-body ground states as Laughlin-like FCIs.

\emph{Quantum geometry perspective}---%
An efficient tool to understand the emergence of FCIs in the moiré trivial band is through the lens of quantum geometry~\cite{PhysRevLett.124.106803,PhysRevResearch.5.L012015,Hui_broken_symmetry,jie_wang_qgt1,jie_wang_qgt2,bo2024}, which refers to the description of the band's eigenstate manifold in terms of Berry curvature and FS metric. Importantly, the latter governs wavefunction overlaps and, thus, emerges in the form factors of the interaction matrix elements. The inhomogeneous distribution of the FS metric in the mBZ breaks the particle-hole symmetry of the band-projected many-body Hamiltonian~\cite{PhysRevLett.111.126802}, leading to a non-constant single-hole energy $E_h({\mathbf{k}})\simeq \sum_{\mathbf{q}}V(\mathbf{q})\exp[{-\sum_{a,b=x,y}q_a q_bg^{ab}(\mathbf{k})}]$ that depends on the FS metric $g^{ab}$~\cite{PhysRevResearch.5.L012015}. In the long-wavelength limit, ${\rm tr} g$ controls $E_h({\mathbf{k}})$,
suggesting that electrons (holes) tend to localize in regions with relatively small (large) ${\rm tr} g$ to minimize the energy. As elucidated below, this mechanism provides a qualitative explanation for the existence of stable FCIs in the $C=0$ band, although the FS metric and the associated single-hole energy $E_h({\mathbf{k}})$ are strictly single-particle indicators. 
A similar phenomenon has also been recently observed in a tight-binding setting~\cite{Lin_zero_chern_FCI}.



To obtain insights based on this framework, we calculate the distribution of ${\rm tr} g$ of the targeted trivial band across the mBZ.
As shown in Fig.~\ref{fig:2}(a), the weak intervalley coupling causes the FS metric to be strongly pronounced and concentrated near the $\kappa$ point, while in the rest of the mBZ it exhibits a nearly uniform distribution (for comparison, the FS metric in unperturbed TBG is shown in the SM~\cite{SupMat}). For finite systems, whether the $\kappa$ point is included in the set of accessible discrete momenta depends on the geometry of the sample. When the $\kappa$ point is accessible, $E_h(\mathbf{k})$ expels electrons away from the region of large FS metric toward regions of small and uniform FS metric, leaving behind an effective ``hole'' at the $\kappa$ point [see Fig.~\ref{fig:2}(d)]. Crucially, the distribution of Berry curvature is very similar to that of the FS metric in our trivial-band model [Figs.~\ref{fig:2}(a) and (b)], so the expelled electrons primarily feel the topological information encoded in the unperturbed TBG, i.e. the nearly uniform Berry curvature across most of the mBZ. In this case, FCIs are favored, as confirmed by the finite many-body energy gap [Fig.~\ref{fig:1}(d)] and the nearly uniform electron density away from $\kappa$ [Fig.~\ref{fig:2}(d)]. The smooth many-body Berry curvature, quantized $\sigma_{xy}$, and PES counting are displayed in the SM~\cite{SupMat}. From this perspective, the nonzero $\sigma_{xy}$ in the trivial band is also naturally explained: the cancellation of opposite topological charges---which would lead to a trivial Hall conductance of the fully occupied mBZ---does not occur in the many-body ground state at $\nu=1/3$. For systems lacking any particular spatial symmetry, the $\kappa$ point is excluded from the accessible momenta. Compared to the $\kappa$-accessible case, electrons are less affected by the peaks of FS metric and Berry curvature in the continuum limit, therefore signatures of FCIs may even be enhanced---we indeed find larger many-body energy gaps and PES gaps in this case [Fig.~\ref{fig:1}(c) and (d)].

While only one or two accessible momentum points reside within the range of the FS metric/Berry curvature peak for the systems reachable by exact diagonalization, we expect the mechanism described above to still work in extremely large systems: electrons will be significantly depleted from all momentum points within the range of the FS metric peak. Since the Berry curvature peak is located at the same place, electrons cannot efficiently feel it even if it is finely resolved by the dense grid of accessible momenta. The FS metric and Berry curvature peaks are rather sharp, so there is sufficient space in the mBZ to host the expelled electrons at a low filling factor such as $\nu=1/3$.

We have also considered the interaction-induced band mixing by involving both the $C=0$ band and the next higher band. In this case, we find that the FCI ground states remain stable (see SM~\cite{SupMat}). 
In contrast to FCIs in tMoTe$_2$~\cite{abouelkomsan2024band, xu2024maximally, yu2024fractional}, band mixing here does not seem to have a significant influence on the many-body energy gap.

\emph{Laughlin-like FCIs in higher-Chern bands}---%
Building on these insights, we now extend the discussion to a more general setting where the single-particle band carries an arbitrary Chern number. 
As a concrete and experimentally relevant example, we focus on twisted double bilayer graphene (TDBG), consisting of two Bernal-stacked
graphene sheets with a relative twist angle in between. This material has been shown to host a nearly flat conduction band with $C=2$ (shown in SM~\cite{SupMat}) within a realistic parameter range~\cite{ZhaoTDBG,Raul_hall_crystal,Jung_tdbg,lee_tdbg}. 
Here, the Chern number $C=2$ originates from the coupling between valleys of the same chirality in the two untwisted layers, which generates an additional topological charge. 

\begin{figure}[t]
\centering
\includegraphics[width=\linewidth]{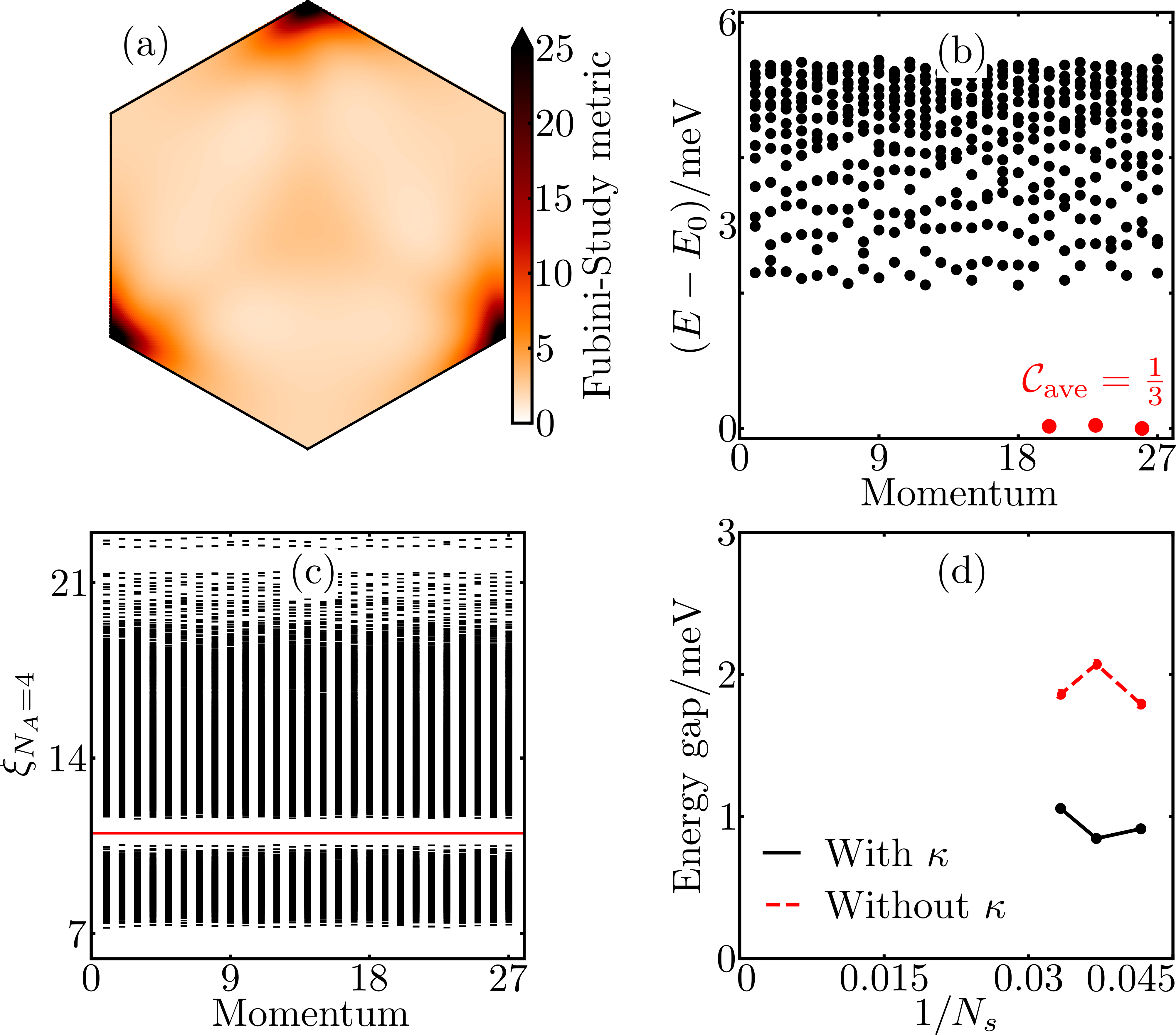}
\caption{\textbf{ Laughlin-like states in a realistic \boldmath{$C=2$} band}. (a) Trace of the FS metric in the mBZ displaying the characteristic concentration at the $\kappa'$ point. (b) Low-lying many-body energy spectrum, (c) PES, and (d) many-body energy gap scaling (for momentum grids with and without the $\kappa$ and $\kappa'$ points) demonstrating the Laughlin-like nature of the many-body ground states at $1/3$ filling of the $C=2$ band.
The data shown in (b) and (c) correspond to a 27-site system 
which excludes the $\kappa$ point from the set of discrete accessible momenta.
The number of states below the solid line in (c) corresponds to the quasihole counting of the $1/3$ Laughlin state. The system parameters are detailed in the End Matter. 
}
\label{fig:3}
\end{figure}

The quantum geometry of this band is qualitatively similar to that of the trivial band studied above. The FS metric and the Berry curvature have similar distributions in the mBZ, both developing a pronounced peak at the $\kappa'$ point, while in the rest of the mBZ exhibiting an almost uniform background; see Fig.~\ref{fig:3}(a) and the SM~\cite{SupMat}.
The same logic as in the $C=0$ case implies the emergence of Laughlin-like FCIs with Hall conductance $\sigma_{xy}=1/3$ rather than $\sigma_{xy}=C\nu=2/3$~\cite{Gunnar2015}. 
From the quantum geometry perspective, the absence of electron occupation at the mBZ corner removes the corresponding single-particle topological information from the many-body state. 
In line with this, we observe compelling evidence of Laughlin-like FCIs in the $C=2$ band with strongly inhomogeneous quantum geometry. The low-energy spectrum with three-fold ground-state degeneracy and the PES with the Laughlin-type counting are shown in Fig.~\ref{fig:3}(b)-(d). In particular, the numerically obtained many-body Chern number of $1/3$ is in clear departure from the conventional $2/3$ expectation (see SM for more details~\cite{SupMat}).

\begin{figure}[t]
\centering
\includegraphics[width=\linewidth]{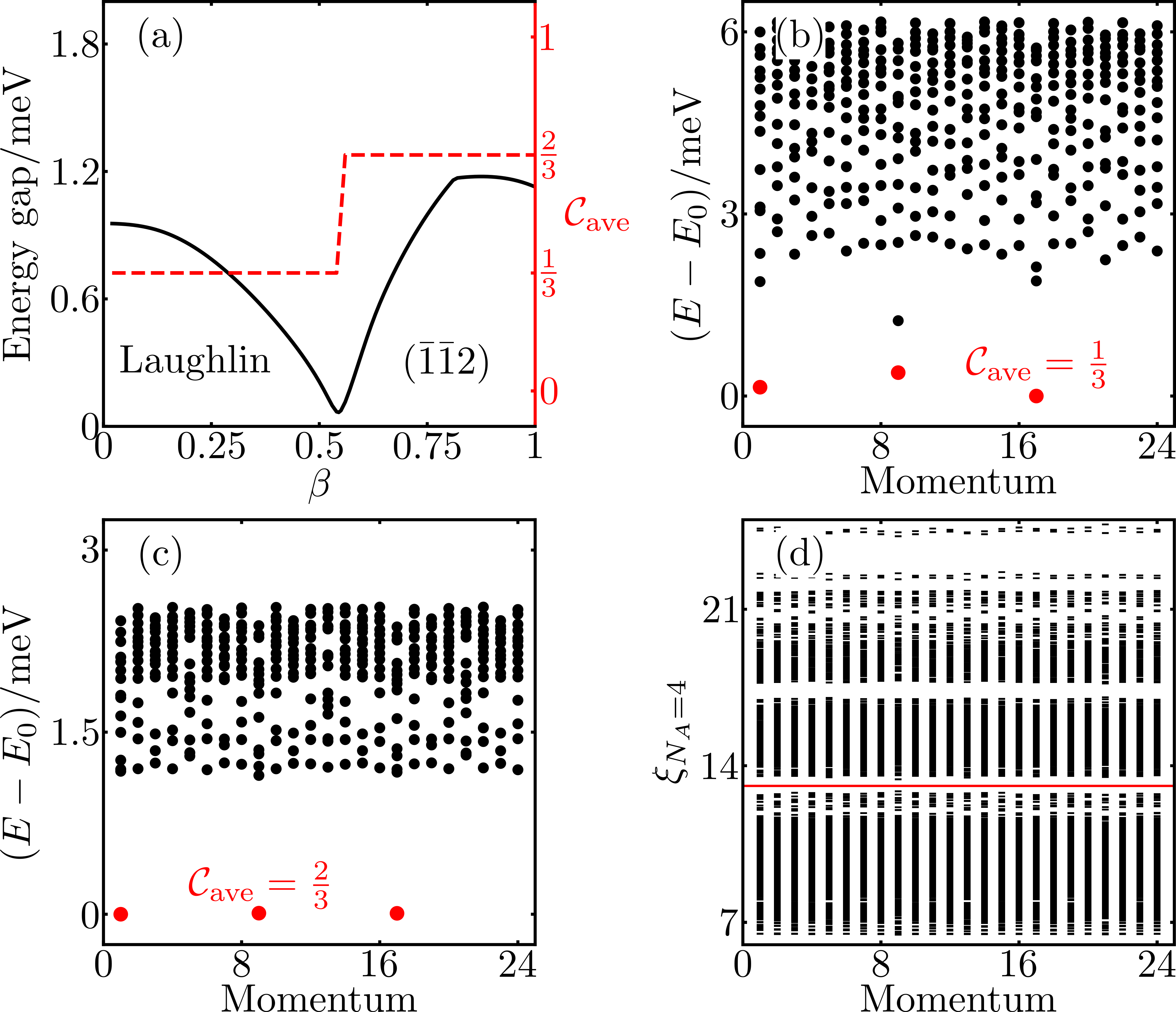}
\caption{\textbf{Phase transition in an ideal \boldmath{$C=2$} band}. (a) Many-body energy gap (black solid line) and many-body Chern number (red dashed line) as a function of interlayer coupling, displaying a clear phase transition. (b) The low-lying many-body energy spectrum at $\beta=0.1$ shows a Laughlin-like threefold degenerate ground state with $\mathcal{C}_{\text{ave}}=1/3$. (c) Low-lying many-body energy spectrum at $\beta=1$ characterized by a threefold degenerate $\mathcal{C}_{\text{ave}}=2/3$ ground state. (d) The corresponding PES for the $\mathcal{C}_{\text{ave}}=2/3$ state contains $5508$ below the solid red line, matching the quasihole state counting of the quantum Hall bilayer $(\bar{1}\bar{1}2)$ state. Here, we considered a $24$-site system which includes the $\kappa$ point. 
}
\label{fig:4}
\end{figure}

\emph{Extension to ideal bands}---%
The stabilization of Laughlin-like FCI states can be extended to ideal flat bands with arbitrary Chern number~\cite{jie_wang_qgt1,PhysRevLett.128.176404}.
To exemplify this, we consider the chiral limit of TDBG at the magic twist angle, possessing an exactly flat $C=2$ conduction band with ideal quantum geometry, i.e., the FS metric and the Berry curvature vary in sync in the mBZ~\cite{jie_wang_qgt2}. At $1/3$ filling, by varying the interlayer coupling between untwisted layers, $\beta$, we observe that the low-lying many-body energy spectrum undergoes a gap closing and reopening near $\beta\approx 0.55$ [see Fig.~\ref{fig:4}(a)], signaling a quantum phase transition. 
Furthermore, our calculations reveal that the Chern number of the many-body ground state changes from $\mathcal{C}_{\text{ave}}=1/3$ to $\mathcal{C}_{\text{ave}}=2/3$, which has been also observed in non-ideal flat bands~\cite{song2025fractionalcherninsulatorstransition}.



Following the same logic as before, we can understand this phase transition from the perspective of the quantum geometry distribution in the mBZ. 
The Chern number $C=2$ of the ideal band in the chiral-magic TDBG originates from the ideal $C=1$ band in the chiral-magic TBG coupled with a Dirac band of the same chirality~\cite{jie_wang_qgt1}. 
Both the FS metric and Berry curvature are strongly peaked at $\kappa$ for small $\beta$ and become flatter at moderately increasing values of $\beta$~\cite{PhysRevLett.128.176404}. Therefore, we expect the $\mathcal{C}_{\text{ave}}=1/3$ Laughlin FCI at small $\beta$, where electrons do not feel the Berry curvature peak. Indeed, this phase exhibits the characteristic features of the Laughlin states, as shown in Fig.~\ref{fig:4}(b) (in addition, the PES is shown in the SM~\cite{SupMat}).
By contrast, the $\mathcal{C}_{\text{ave}}=2/3$ phase should be an authentic $C=2$ FCI, in which electrons feel the entire Berry curvature of the $C=2$ band. In this case, although the ground states may appear in the same momenta as the Laughlin FCIs, the PES shows different quasihole counting. In fact, the PES counting matches an unconventional quantum Hall bilayer $(\bar1 \bar12)$ state, where each layer forms an integer quantum Hall effect of composite fermions under magnetic field with opposite direction~\cite{PhysRevB.53.15845}.
The emergence of Laughlin and Halperin states in a $C=2$ band has also been predicted in graphene under periodic strain modulation~\cite{Zheng_periodicstrain}.

\emph{Conclusion and discussion}---%
Based on an exact diagonalization approach, we have demonstrated that fractional Chern insulators can emerge in a topologically trivial moiré band under long-range Coulomb interactions---establishing the presence of FCI physics in realistic systems far beyond the Landau-level paradigm. 
We highlight the distribution of quantum geometry across the Brillouin zone---and not the single-particle Chern number---as the key heuristic principle driving the stabilization of FCIs in flat bands.
In the specific system considered here, electrons follow a nearly homogeneous distribution across most of the mBZ---avoiding the $\kappa$ point, where the FS metric is highly concentrated---and experience a nearly-constant Berry curvature which facilitates the formation of Laughlin-like many-body states.
We have further showed that this quantum geometric mechanism extends to arbitrary Chern numbers in both realistic and ideal twisted multilayer graphene systems. Here, the mismatch between Hall conductance and the conventional expectation underscores the limitations of standard topological band arguments and points to richer phase diagrams in higher-Chern bands.
We expect that future theoretical work will formalize the generic conditions for stabilizing Laughlin states purely in terms of quantum geometry and independently of the band Chern number.

Besides unveiling a geometric heuristic principle for the stabilization of FCIs, our work provides a recipe for most strikingly showcasing these physics. In particular, we have showed that coupling two valleys of opposite chirality leads to a trivial band---which would otherwise have a $|C|=1$ Chern number---hosting Laughlin-like FCI ground states at fractional filling.
As a specific experimentally relevant example of this mechanism, trivial moiré bands hosting FCIs can be engineered in twisted bilayer graphene via intervalley coupling induced by a commensurate substrate~\cite{intervalley_coupling1,intervalley_coupling}.
In addition, our results indicate that multilayer systems hosting higher-Chern bands constitute a promising platform for investigating the competition between unconventional FCIs and Halperin-like states~\cite{PhysRevLett.52.1583}, where the combination of quantum geometry and single-particle layer polarization---rather than the band topology---could play a dominant role.

The unveiled quantum-geometric mechanism is not only crucial for stabilizing Laughlin states in a generic setting beyond ideal Chern bands and Landau levels, but it could also prove relevant for a deeper understanding of more exotic (e.g. non-Abelian) topological order.
Moreover, it suggests new pathways for designing exotic quantum matter with mismatched Hall conductance.
In particular, engineering the quantum geometry distribution could serve as a heuristic approach for stabilizing sought-after fractional Hall crystals~\cite{Maria2014,Song2023,hongyu2025} and recently introduced anti-topological phases~\cite{reddy2024antitopologicalcrystalnonabelianliquid}. Related phases exhibiting a quantized Hall conductance differing from the filling factor have been predicted in lattices, employing a composite fermion picture based on Chern-Simons theory~\cite{Kol1993,Santos2018,Santos2024}. Exploring how these different perspectives intertwine will be a rewarding venture which will certainly improve our understanding of correlated topological phases.

\emph{Data Availability}---%
The data that support the findings of this article are not publicly available. The data are available from the authors upon reasonable request.

\emph{Acknowledgments}---%
We acknowledge useful discussions with Hongyu Lu. H.L., R.P.C., and E.J.B. were supported by the grants awarded to E.J.B. by the Swedish Research Council (2018-00313 and 2024-04567), the Knut and Alice Wallenberg Foundation (2018.0460 and 2023.0256), and the Göran Gustafsson Foundation for Research in Natural Sciences and Medicine. Z. L. was supported by the National Natural Science Foundation of China (Grants No. 12350403 and
No. 12374149). 
R.P.C. acknowledges additional funding from the Swedish Royal Academy of Sciences foundations (Kungl. Vetenskapsakademierna stiftelser) and from the Wenner-Gren foundations.
In addition, we utilized the Sunrise HPC facility supported by the Technical Division of the Department of Physics, Stockholm University.

\bibliography{reference}

\onecolumngrid

\section{Supplemental Material for ``Topological Order Without Band Topology in Moiré Graphene''}

In this supplementary material, we provide additional results for the trivial band, the realistic and ideal $C=2$ band systems, as well as information for tilted samples.

\section{Model setting}%
In the main text, we have shown the results with many-body interactions. Here, we provide the detailed single-particle information.
For the $C=0$ band, we use a twisted bilayer graphene, with the bottom layer having an intervalley coupling induced by an insulating substrate of commensurate lattice~\cite{intervalley_coupling1}. 
The top layer is modeled by the Hamiltonian
\begin{eqnarray}
    H^{t}_1=\begin{pmatrix}
        U&\frac{\sqrt{3}}{2}at_0(k_x^{t}-ik_y^{t})\\
        \frac{\sqrt{3}}{2}at_0(k_x^{t}+ik_y^{t})&-U
    \end{pmatrix}.
\end{eqnarray}
The bottom layer with a weak intervalley coupling leads to 
\begin{eqnarray}
    H^{b}_1=\begin{pmatrix}
        U&\frac{\sqrt{3}}{2}at_0(k_x^{b}-ik_y^{b})&0&t_1\\
        \frac{\sqrt{3}}{2}at_0(k_x^{b}+ik_y^{b})&-U&0&0\\
        0&0&U&\frac{\sqrt{3}}{2}at_0(k_x^{b}+ik_y^{b})\\
        t_1&0&\frac{\sqrt{3}}{2}at_0(k_x^{b}-ik_y^{b})&-U
    \end{pmatrix}.
\end{eqnarray}
In addition, the two layers are coupled by a moiré periodic potential, $T_j=w_0\sigma_0+w_1e^{i(2\pi/3)j\sigma_z}\sigma_xe^{-i(2\pi/3)j\sigma_z}$, which connects the same valleys of different layers. Here, $a=2.46\,\text{\AA}$ is the lattice constant of graphene and $t_0=2610$meV is the nearest-neighbor hopping amplitude. With $\theta=1.13^\circ$, $(w_0, w_1, U,t_1)=(30, 110, 10, 20)$meV, one obtains the single-particle dispersion in the main text, where the flat band above charge neutrality governs a Chern number $C=0$.  
Here, the inclusion of intervalley coupling mainly induces a concentrated Berry curvature near one of the inequivalent corners of mBZ, while the remainder of the zone remains nearly uniform.  
For a more realistic setting, it would be important to include the effect of both valleys of the top layers, as well as the spin degree of freedom~\cite{intervalley_coupling}. However, since our main focus is on the disconnect between many-body topological states and single-particle band topology, we defer such an extension to future work.

\begin{figure}[t]
\centering
\includegraphics[width=0.5\linewidth]{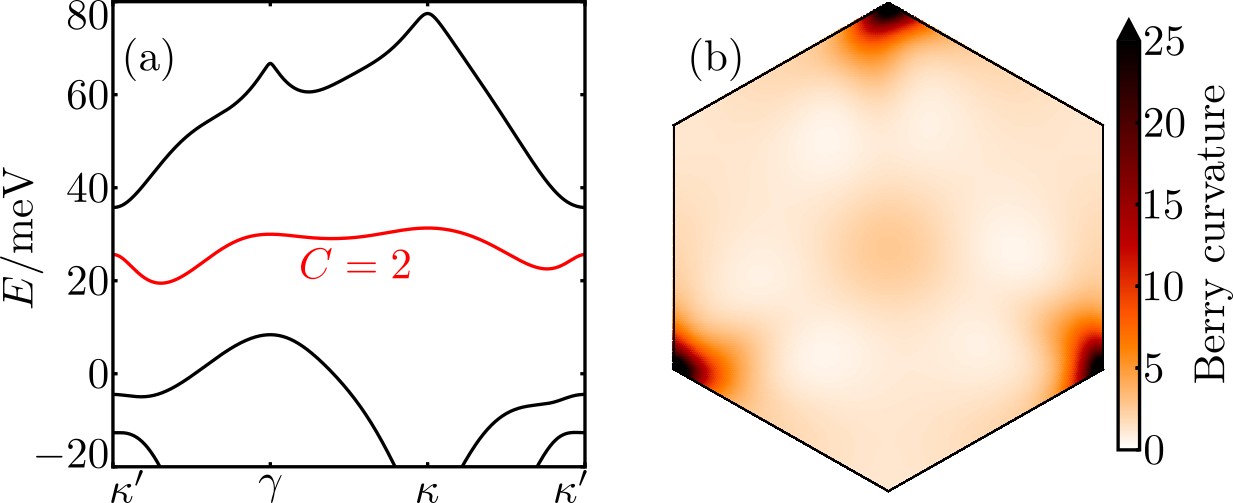}
\caption{ The targeted band of TDBG above charge neutrality with $C=2$ (a) and the corresponding Berry curvature distribution over mBZ. 
}
\label{fig:6}
\end{figure}

For the $C=2$ band, we consider two sheets of AB stacked bilayer graphene, twisted relative to each other by a small angle. Each sheet is described by the single-particle Hamiltonian~\cite{ZhaoTDBG,Raul_hall_crystal,Jung_tdbg,lee_tdbg}
\begin{eqnarray}
    H^{t,b}_2=\begin{pmatrix}
        U_1+\delta&\frac{\sqrt{3}}{2}at_0(k_x^{t,b}-ik_y^{t,b})&-\frac{\sqrt{3}}{2}at_4(k_x^{t,b}+ik_y^{t,b})&t_1\\
        \frac{\sqrt{3}}{2}at_0(k_x^{t,b}+ik_y^{t,b})&U_1&-\frac{\sqrt{3}}{2}at_3(k_x^{t,b}-ik_y^{t,b})&-\frac{\sqrt{3}}{2}at_4(k_x^{t,b}+ik_y^{t,b})\\
        -\frac{\sqrt{3}}{2}at_4(k_x^{t,b}-ik_y^{t,b})&-\frac{\sqrt{3}}{2}at_3(k_x^{t,b}+ik_y^{t,b})&U_2&-\frac{\sqrt{3}}{2}at_0(k_x^{t,b}-ik_y^{t,b})\\
        t_1&-\frac{\sqrt{3}}{2}at_4(k_x^{t,b}-ik_y^{t,b})&-\frac{\sqrt{3}}{2}at_0(k_x^{t,b}+ik_y^{t,b})&U_2+\delta
    \end{pmatrix}.
\end{eqnarray}
The two sheets $H_2^{t}$ and $H_2^{b}$ are also coupled through the moiré periodic potential. 
For realistic TDBG, we choose $(w_0,w_1,t_0,t_1,t_3,t_4,\delta)=(70, 130, 2610, 361, 283, 138, 15)$meV and use $(U_1,U_2)=(30,10)$meV for the top layer and $(U_1,U_2)=(-10,-30)$meV for the bottom layer. The band dispersion and Berry curvature distribution at $\theta=1.35^\circ$ are displayed in Fig.~\ref{fig:6}.
Furthermore, by setting $w_0=t_3=t_4=\delta=U_1=U_2=0$, this realistic TDBG model reduces to the chiral limit. The magic angle is determined by $\sin(\theta/2)=\frac{\sqrt{3}w_1}{4\pi\alpha t_0}$ with $\alpha\approx 0.586$. We choose $w_1=110$meV for the chiral-magic TDBG which supports exact flat ideal bands. $t_1$, which is converted to a dimensionless parameter $\beta=t_1\alpha/w_1$ in the section of ideal $C=2$ bands, is tuned to drive the many-body phase transition.

\section{Trivial band}

\begin{figure}[h]
\centering
\includegraphics[width=0.95\linewidth]{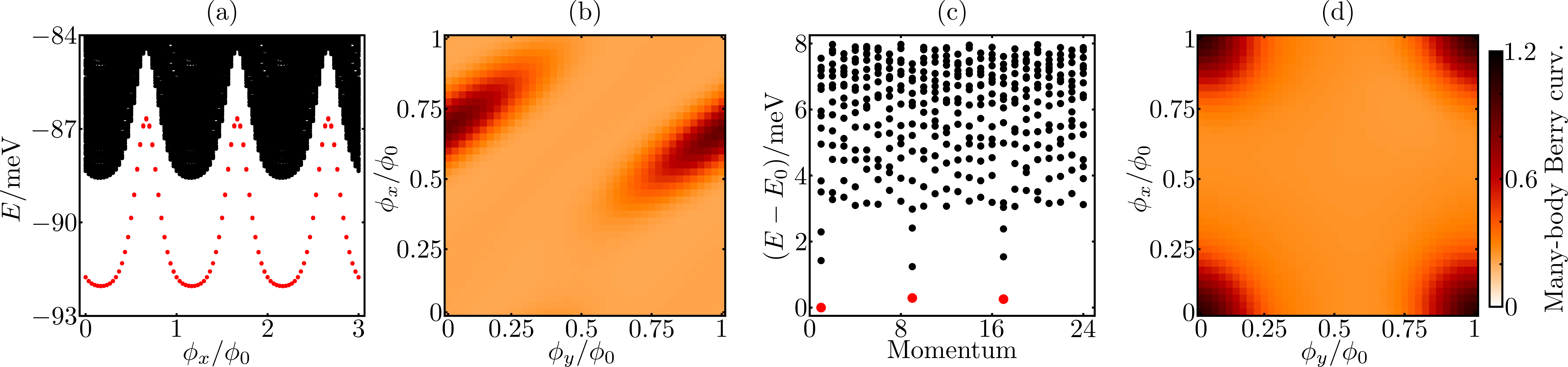}
\caption{Spectral flow (a) and averaged many-body Berry curvature (b) of ground states of the $27$-site cluster (where $\kappa$ is part of the momentum grid) for the trivial band system. (c) and (d) are the energy spectrum and the averaged many-body Berry curvature of the $24$-site cluster (where $\kappa$ is not part of the momentum grid), respectively. The Berry curvature $\Omega(\phi_x,\phi_y)$ is normalized by the number of flux points $N_\phi=40^2$, i.e., $\Omega(\phi_x,\phi_y) N_\phi/2\pi$. In energy spectrum plots, the red dots indicate the three Laughlin-like ground states.
}
\label{fig:c_0_spectrum_many_body_berry}
\end{figure}

\begin{figure}[h]
\centering
\includegraphics[width=0.45\linewidth]{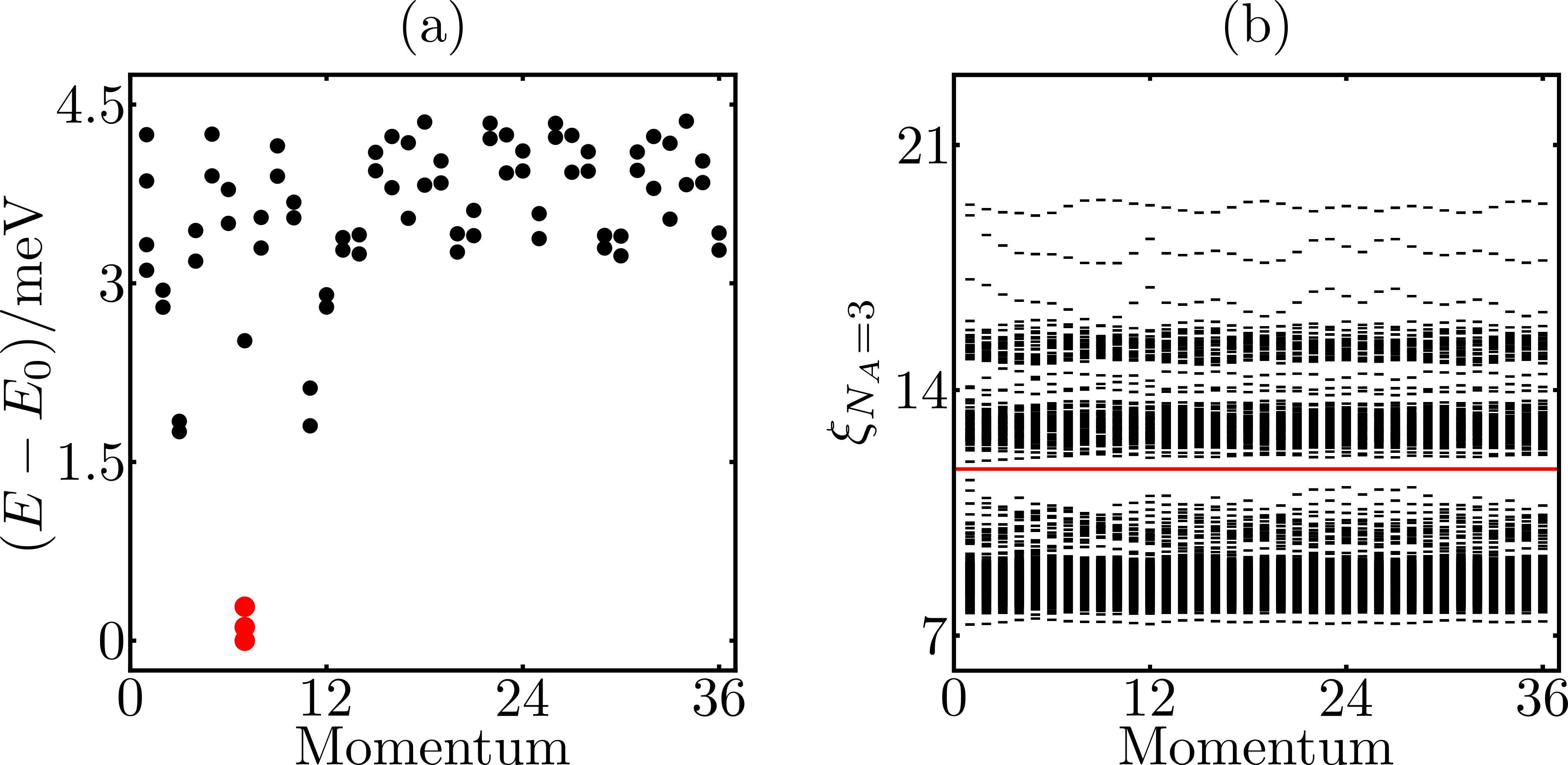}
\caption{Low-lying energy spectrum (a) and particle-cut entanglement spectrum (b) of the $36$-site cluster for the trivial band system, respectively. The number of states below the first entanglement gap is $4872$, corresponding to the quasihole counting of Laughlin-like states. Here, the considered momentum grid contains the $\kappa$ point.
}
\label{fig:36site}
\end{figure}

In this section, we provide additional results for the trivial band system. 
Having shown the low-lying energy spectrum and particle-cut entanglement spectrum of the $27$-site cluster (excluding the $\kappa$ point) in the main text, we now include the corresponding spectral flow and many-body Berry curvature distribution in Fig.~\ref{fig:c_0_spectrum_many_body_berry}(a) and (b). 
Both indicate a smooth evolution as a function of external magnetic flux with no gap closings. Integrating the many-body Berry curvature gives a many-body Chern number $\mathcal{C}_{\text{ave}}=1/3$.
We note that the increase of the ground state energy above the energy of excited states at a different flux in Fig.~\ref{fig:c_0_spectrum_many_body_berry}(a) is not physically indicative of a low stability of the FCI.
We next consider the $24$-site cluster, now including the $\kappa$ point. In this case, the energy spectrum still exhibits three Laughlin-like ground states, which again share an averaged many-body Chern number $1/3$ (see Fig.~\ref{fig:c_0_spectrum_many_body_berry}(c) and (d)). Note that the energy gap decreases when including the $\kappa$ point, potentially signaling a competition with commensurate CDW order. Importantly, the ground-state degeneracy, entanglement gap, and many-body Chern number remain consistent with Laughlin states.
Similar behavior is observed in larger systems. 
As shown in Figs.~\ref{fig:27site_kappa} and \ref{fig:36site} for 27- and 36-site clusters that contain $\kappa$, the energy spectrum displays three degenerate Laughlin-like states (with $\mathcal{C}_\text{ave}=1/3$) in the correct momentum sectors separated from excited states by a clear gap. 
The particle-cut entanglement spectrum with $N_A=3$ reproduces the quasihole counting expected for Laughlin-like states, although with a small entanglement gap in the case of the 27-site cluster.
These results demonstrate that our findings are consistent and robust across different system sizes (both including and excluding the $\kappa$ point).

\begin{figure}[h]
\centering
\includegraphics[width=0.45\linewidth]{27site_with_kappa.png}
\caption{(a) Low-lying energy spectrum and (b) spectral flow at filling $\nu=1/3$ for a $27$-site cluster including the $\kappa$ point. The red dots indicate the threefold degenerate ground states with an average many-body Chern number $\mathcal{C}_{\text{ave}}=1/3$. (c) and (d) are the many-body Berry curvature distribution and the particle-cut entanglement spectrum for $N_A=4$ of the ground states. In panel d, the number of states below the red line corresponds to the quasihole statistics of Laughlin states.
}
\label{fig:27site_kappa}
\end{figure}

\begin{figure}[h]
\centering
\includegraphics[width=0.46\linewidth]{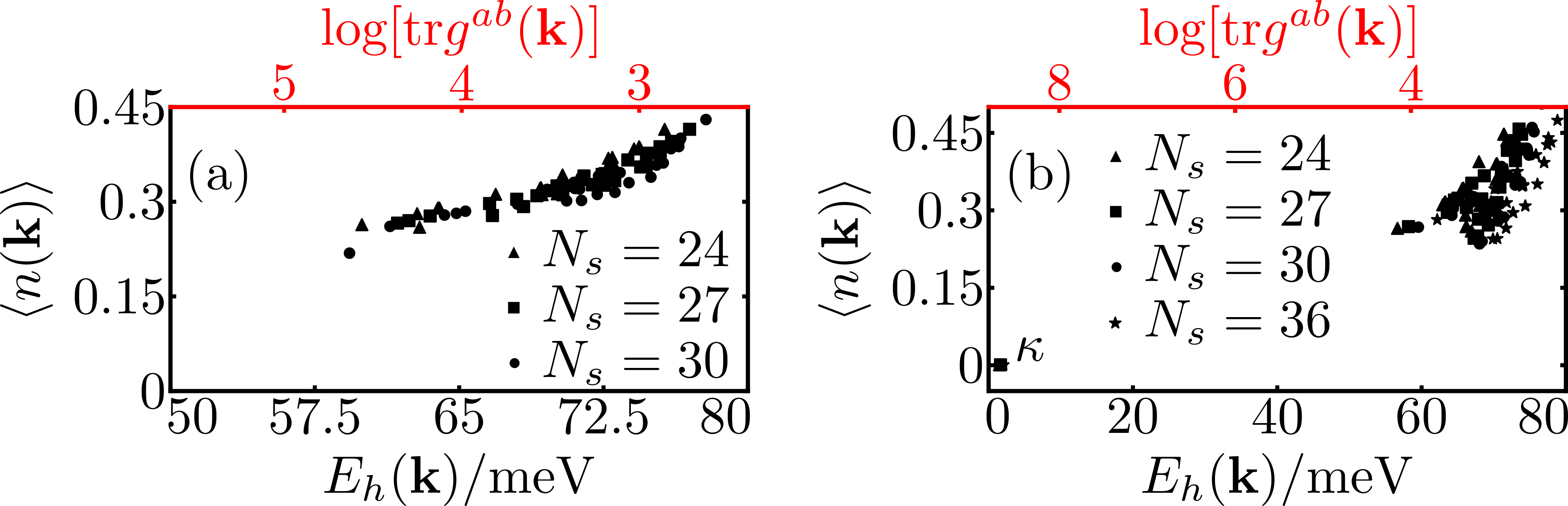}
\caption{Electron occupation as a function of single-hole energy of different samples without $\kappa$ (a) and with $\kappa$ points (b) for the trivial band system, respectively.
}
\label{fig:nk_hole}
\end{figure}

In the main text, we have shown that the electron occupation is organized by quantum geometry.
Here, we examine the relation between electron occupation and single-hole energy across various system sizes and shapes. 
In Fig.~\ref{fig:nk_hole}(a) and (b), as expected by the quantum geometry perspective, electrons occupy states with higher single-hole energy, while states with low single-hole energy (i.e., near the $\kappa$ point) tend to remain empty. 
Consequently, the electron occupation is roughly uniform at $1/3$ away from $\kappa$ and nearly zero at $\kappa$---a behaviour that appears consistently across various system sizes.

We note that, while the sparse momentum sampling in the considered finite clusters does not allow to resolve multiple points of the Berry curvature peak directly, the effect of this peak is captured in systems that include the $\kappa$ point and also emerges by flux insertion. The effect becomes evident in the reduction of the gap and the absent $\kappa$-point occupation in the former case, and in the pronounced variation of energy and many-body Berry curvature in the latter case.
The persistence of the many-body energy gap for different systems including and excluding the $\kappa$ point, together with the consistent $\mathcal{C}=1/3$ Chern number for dense flux grids, strongly support the stabilization of a trivial-band FCI in the thermodynamic limit.

\begin{figure}[tb]
\centering
\includegraphics[width=0.47\linewidth]{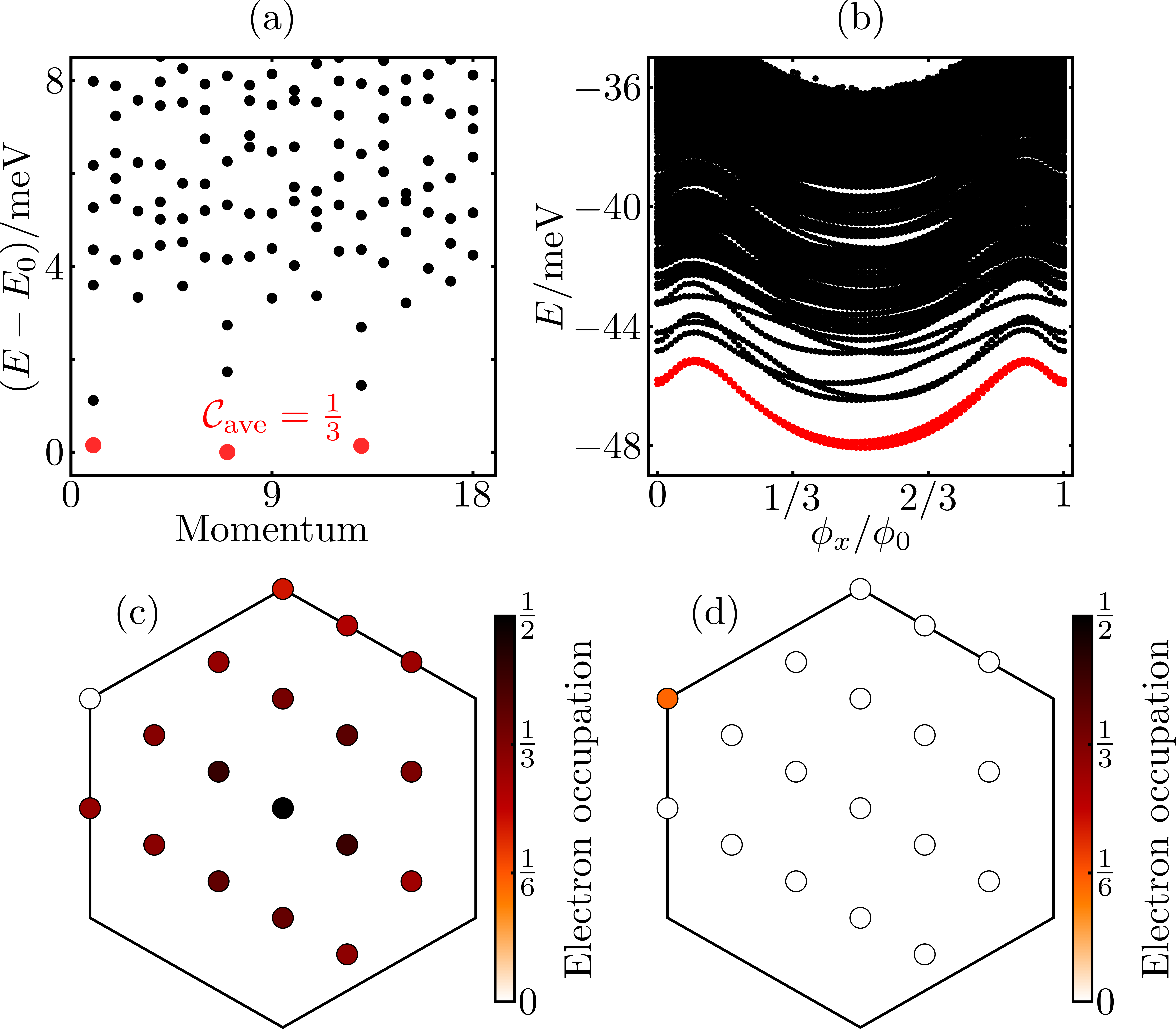}
\caption{(a) and (b) Low-lying energy spectrum and spectral flow at $1/3$ electron filling obtained by two-band exact diagonalization. The red dots indicate the threefold degenerate ground states with an average many-body Chern number $\mathcal{C}_{\text{ave}}=1/3$. (c) and (d) are the electron occupation for the trivial band and for the higher band. Here, the tilted sample has a spanning vector $\mathbf{T}_1=(4, 2)$ and $\mathbf{T}_2=(1, 5)$ to include the $\kappa$ point.
}
\label{fig:2bandED13}
\end{figure}

In addition, we have performed two-band exact diagonalization to ensure that the FCI in the trivial band is not destroyed by band mixing.
The obtained many-body energy spectrum, many-body Chern number, and spectral flow (see Fig. \ref{fig:2bandED13}) show that the FCI persists when considering the impact of the higher band. Again, we observe that the electron occupation in the lowest band avoids $\kappa$, although now there is a sizable occupation at this point in the higher band. Thus, now the overall occupation is more homogeneous, as expected from an incompressible liquid. Remarkably, the energy gap remains mostly unchanged by the inclusion of the second band.

\begin{figure}[h]
\centering
\includegraphics[width=0.5\linewidth]{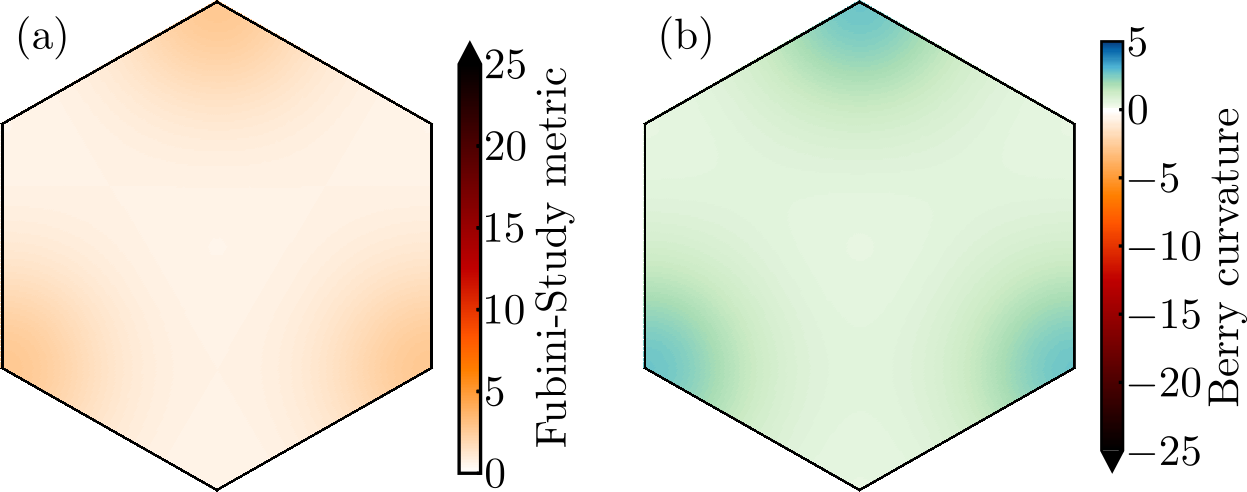}
\caption{Fubini-Study metric (a) and Berry curvature (b) distribution over mBZ for the band without intervalley coupling. For ease of comparison, the same colorbar as in the main text is used.
}
\label{fig:c_1_qm_bc}
\end{figure}

Finally, we provide more details of the target band before introducing intervalley coupling. 
In this case, the Berry curvature and quantum metric are uniform, supporting a Chern number $C=1$ (see Fig.~\ref{fig:c_1_qm_bc}). 
Comparing this to the Berry curvature and quantum metric distribution after intervalley coupling is incorporated (shown in the main text), it is clear that intervalley coupling introduces an opposite topological contribution at the $\kappa$ point, resulting in a trivial band.

\section{results for the realistic $C=2$ band} 

\begin{figure}[t]
\centering
\includegraphics[width=0.45\linewidth]{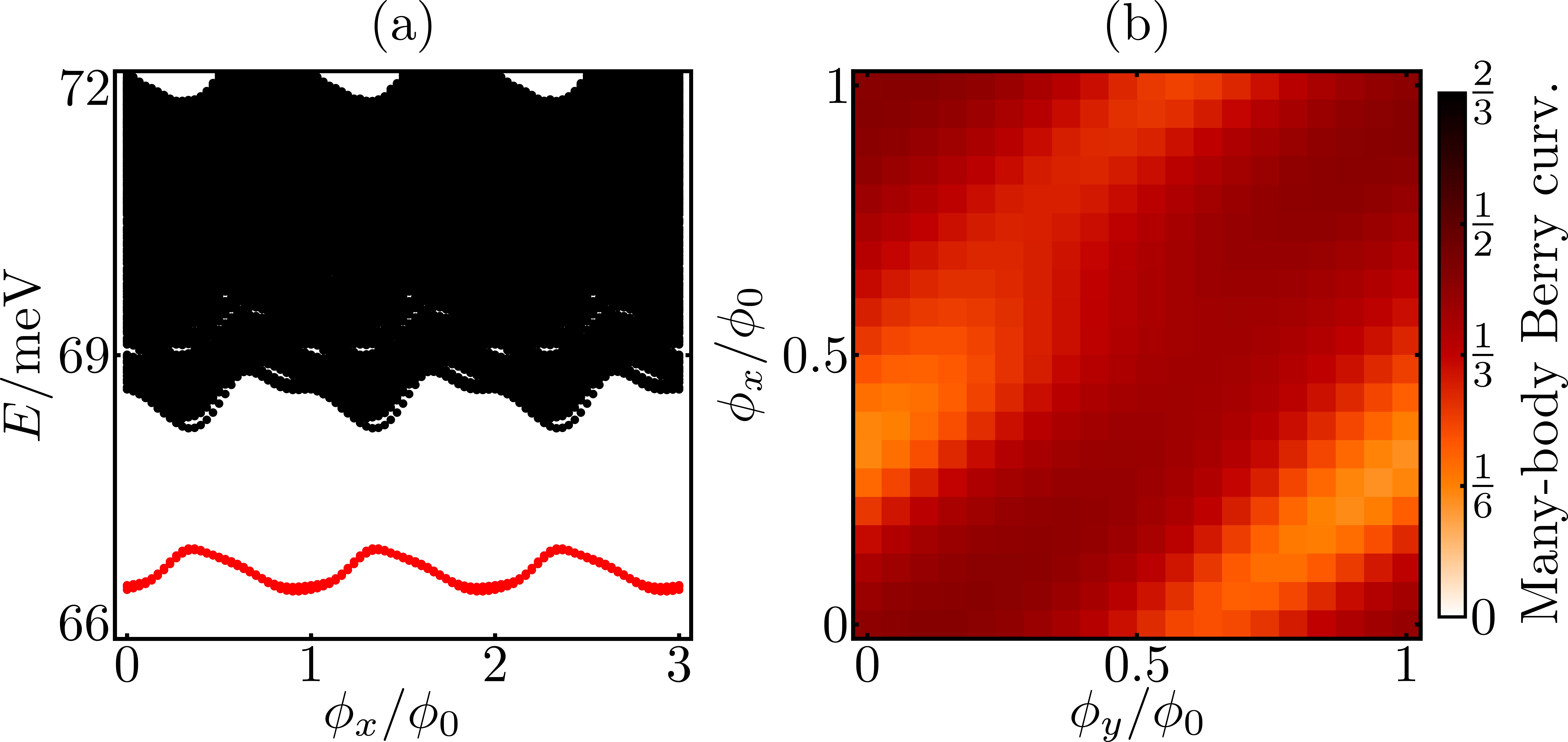}
\caption{Spectral flow (a) and many-body Berry curvature distribution (b) of Laughlin-like states for the realistic $C=2$ band.
}
\label{fig:c_2berry}
\end{figure}

In the main text, we have presented the energy spectrum and the particle-cut entanglement spectrum for the $27$-site cluster of the realistic $C=2$ band. 
Here, we provide the corresponding spectral flow and the many-body Berry curvature, shown in Fig.~\ref{fig:c_2berry}(a) and (b). 
In both cases, the gap remains open under the insertion of an external magnetic flux, and the nearly uniform many-body Berry curvature leads to a quantized Hall conductance $\sigma_{xy}=1/3$.

\begin{figure}[t]
\centering
\includegraphics[width=0.55\linewidth]{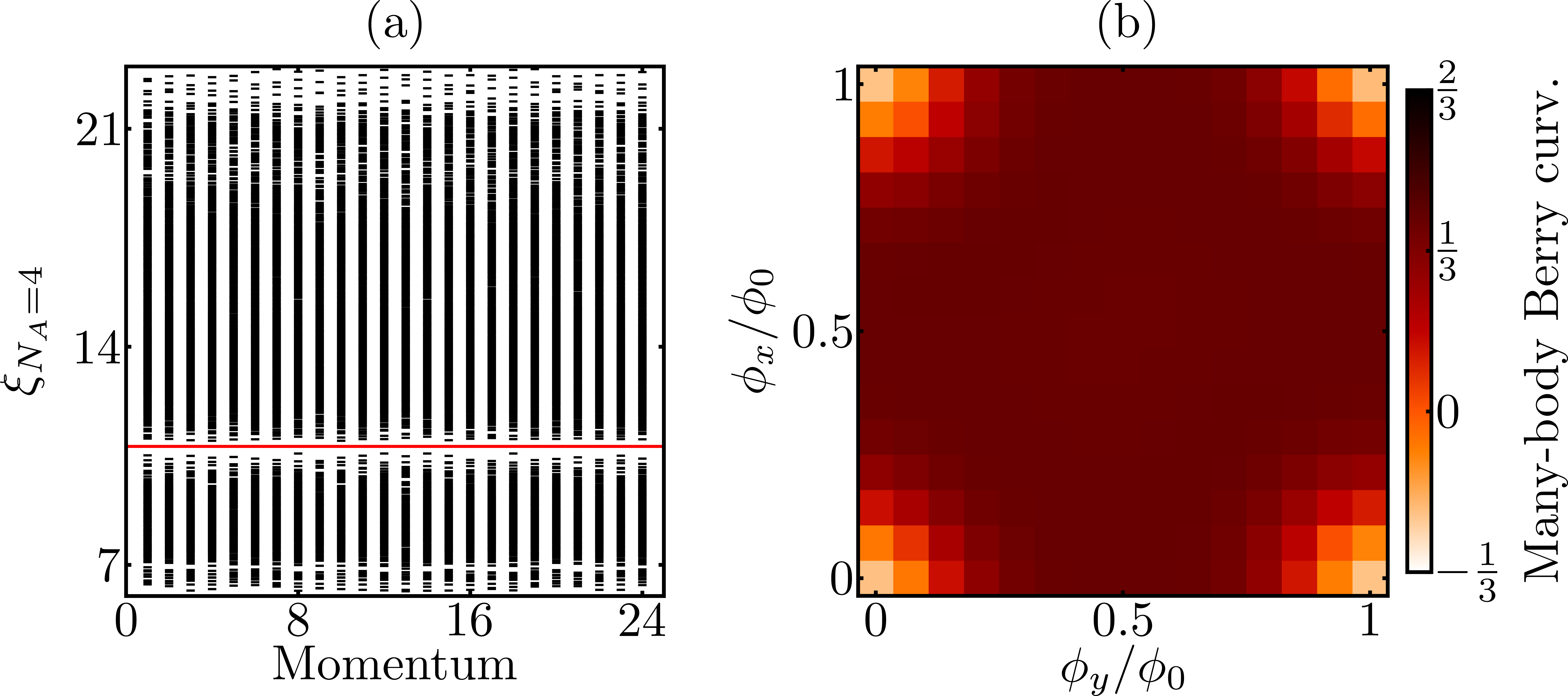}
\caption{Particle-cut entanglement spectrum (a) and many-body Berry curvature distribution (b) of the $24$-site cluster for the ideal $C=2$ band, respectively. The number of states below the first entanglement gap is $2730$, corresponding to the quasihole counting of Laughlin-like states.
}
\label{fig:ideal_c_2}
\end{figure}

\section{results for the ideal $C=2$ band} 
In the main text, we have studied the phase transition between Laughlin-like states and $\bar{1}\bar{1}2$ states in the ideal $C=2$ band and shown data for $\bar{1}\bar{1}2$ states. 
Here, we present additional data for the Laughlin-like states. 
Fig.~\ref{fig:ideal_c_2}(a) displays the particle-cut entanglement spectrum of Laughlin-like states in the ideal higher-Chern band. 
Although the first entanglement gap is small, the number of states below it remains consistent with the exclusion rule of Laughlin-like states. Correspondingly, the smooth many-body Berry curvature yields a many-body Chern number $\mathcal{C}_{\text{ave}}=1/3$. 

\section{tilted samples}
In the work, we employ tilted samples to include or exclude the high symmetry point~\cite{PhysRevB.90.245401}, thereby demonstrating the robustness of our results. 
For the $24$-site cluster, we use the spanning vectors $\mathbf{T}_1=(5, 1)$ and $\mathbf{T}_2=(1, 5)$, as well as $\mathbf{T}_1=(4, 0)$ and $\mathbf{T}_2=(0, 6)$.
For the $27$-site cluster, we employ $\mathbf{T}_1=(6, 3)$ and $\mathbf{T}_2=(3, 6)$, $\mathbf{T}_1=(6, 3)$ and $\mathbf{T}_2=(1, 5)$, as well as $\mathbf{T}_1=(6, 1)$ and $\mathbf{T}_2=(9, 6)$.
For the $30$-site cluster, we use $\mathbf{T}_1=(5, 0)$ and $\mathbf{T}_2=(0, 6)$, as well as $\mathbf{T}_1=(4, -1)$ and $\mathbf{T}_2=(6, 6)$.
For the $36$-site cluster, we use $\mathbf{T}_1=(6, 0)$ and $\mathbf{T}_2=(0, 6)$, as well as $\mathbf{T}_1=(6, 0)$ and $\mathbf{T}_2=(3, 6)$.

\end{document}